\documentclass[12pt]{article}
\usepackage{epsfig}

\newcommand{\bn}{\bar\nu}

\newcommand{\sqt}{\sqrt{3}}

\newcommand{\oot}{\overline {126}}

\newcommand{\nnu}{\nonumber\\}
\newcommand{\be}{\begin{equation}}
\newcommand{\ee}{\end{equation}}
\newcommand{\bea}{\begin{eqnarray}}
\newcommand{\eea}{\end{eqnarray}}

\begin{document}

 \vfil
 \vspace{3.5 cm} \Large{
 \title{\bf  {   Fermion Mass Hierarchy  in the Nu MSGUT
 I :  The Real Core   }}
 \author{ Charanjit S. Aulakh   }}
\date{}
\maketitle

 \normalsize\baselineskip=15pt

 {\centerline  {\it
Dept. of Physics, Panjab University}} {\centerline{ \it
{Chandigarh, India 160014}}}

 {\centerline {\rm E-mail: aulakh@pu.ac.in }}
\vspace{1.5 cm}

\large {\centerline{\bf {ABSTRACT }}}
\normalsize\baselineskip=15pt

\vspace{1. cm} The recent  failure of  the scenario which uses
only $\mathbf{10,\oot}$ Higgs multiplets  to yield large enough
 neutrino masses in the $\mathbf{SO(10) }$
 \textbf{MSGUT}  motivates an alternative
scenario\cite{blmdoom} where the  $\mathbf{120}$ -plet collaborates with
the $ \mathbf{10}$ -plet to fit the dominant charged fermion masses. The
small  $\mathbf{\oot} $ -plet couplings   give appreciable contributions
only to light charged fermion masses {\textit{and}} enhance the
Type I seesaw masses  to  viable  values.
  We analyze the 2-3 generation   core of the complete
hierarchical fermion mass system in the CP  conserving
approximation.   \emph{Ansatz} consistency  \emph{requires}
   $  {\mathbf{m_b-m_s = m_{\tau}-m_{\mu}}} $  at the GUT scale $M_X$
    and predicts
near maximal (PMNS) mixing in the leptonic sector for central
values of charged fermion parameters and for   wide ranges of the
other relevant parameters : righthanded neutrino masses and
relative strength of contributions of the two doublet pairs from
the $\mathbf{120}$-plet to  the effective  MSSM Higgs pair. These features
are preserved in the CP preserving 3 generation system  whose
results are previewed : an additional consistency requirement
 $\mathbf{\theta_{13}^c =\theta_{12}^c  \theta_{23}^c }$ on the CKM angles
at $M_X$ arises for 3 generations. Right handed neutrino masses in
such scenarios are all less than about $10^{12} GeV$.

\normalsize\baselineskip=15pt
\maketitle
\section{Introduction}

 Renormalizable supersymmetric $SO(10)$ GUTs\cite{aulmoh,ckn} have
 recently\cite{abmsv}   been a focus of
intense interest both as regards the GUT scale symmetry
breaking\cite{abmsv},
 spectra\cite{ag1,bmsv,fuku04,ag2},RG evolution\cite{ag2} and
 matter Higgs Clebsches \cite{ag1,ag2}.  Even more interest
 arose due to the
demonstration of the feasibility\cite{bsv} of the \emph{generic}
Babu-Mohapatra(BM) program\cite{babmoh} for  a completely
realistic fit of all the fermion mass and mixing data using
\emph{only}
 the $\mathbf{10}$ and $\mathbf{126}$-plet
 Fermion Mass  (FM) Higgs. This scenario
 was initially also considered
to be ``predictive'' in the neutrino sector. Detailed
analysis\cite{gohmoh,bert,babmacesanu} has shown,
 however, that   successful charged fermion
fits in fact require  (arbitrary) choices of the (many) free
phases present in the problem. Thus the successful fits can at
most be considered  indicative  of likely $MSSM\subset  GUT $
embedding angles/phases\cite{gmblm} when the  leptonic mixing and
measured neutrino mass splittings are taken as data. These
embedding angles are physical in the sense that they affect
\cite{gmblm} the baryon violation and Lepton flavour violation
rates in the Susy GUT and may therefore eventually be measurable.
The renewed interest and excitement in the Babu-Mohapatra scenario
began with the demonstration\cite{bsv} that   a natural linkage
existed --at least regarding magnitude if not exact maximality -
between Type II Seesaw\cite{seesaw} dominated neutrino masses and
the manifest approximate  $b-\tau$ unification at scales $\sim
M_{GUT}$ observed in the MSSM and the near maximal neutrino (PMNS)
mixing\cite{strumviss} in the 2-3 sector.
 The nearly successful  generalizations of the generic
 fits to the 3 generation (real) and realistic 3 generation (complex)
 case\cite{gohmoh}  and the ``completely'' satisfactory
generic fits when a small perturbation of the charged fermion masses
 by the $\mathbf{120}$-plet representation was allowed\cite{bert}
 seemed to elevate this generic fitting  scenario to near canonical
  status  : particularly due to the natural
  ease  of the Bajc Senjanovic Vissani   $b-\tau $-Type II seesaw-
  large mixing linkage;
   even though the near exact maximality of the 2-3
  sector leptonic mixing does not have any obvious cause within such
  scenarios. The  early   observation\cite{ckn,gohmohnasri}
  that the Type I and Type II neutrino   masses  tend to
 be too small (and therefore\cite{gohmohnasri})
 may be considered to motivate  more complicated GUT Higgs sectors) did
not dampen the general enthusiasm. However, nearly simultaneously
with the demonstration \cite{bert} that the Type II weak $
\mathbf{{10\oplus\oot \oplus_{weak} 120}}$-plet FM fit was
completely satisfactory at least as far as neutrino mixing
 angles and mass squared splitting  ratios are concerned,
came the surprising finding \cite{babmacesanu} that Type I and
 Type I $\oplus$ Type II fits are -in principle -equally  viable.
Furthermore we showed, nearly simultaneously\cite{gmblm}, that
Type II seesaw fits were most likely always far sub-dominant to
Type I seesaw and Type I itself was too weak in the
BM\cite{babmoh} scenario, at least in the fully specified and
calculable   context of the MSGUT\cite{abmsv,ag2}, and under an
assumption of the genericity of the magnitude of a
 seesaw matrix derived from a successful Type I seesaw fit
 available in the literature\cite{babmacesanu,blmdoom},
\emph{even at the special points where the seesaws are weighted
strongly}\cite{abmsv,bmsv}. Furthermore, with improved
parametrization of the MSGUT symmetry breaking \cite{bmsv2} we
showed\cite{blmdoom},  by a detailed survey of the parameter space
of the MSGUT -assuming the typical values of the
$\mathbf{10-\oot}$ parameters available in the literature
\cite{bert,babmacesanu} were generic -
 that, not only was Type II seesaw subdominant to
Type I Seesaw everywhere in the parameter space,  but also Type I
seesaw itself could not yield neutrino masses in excess of about
$.005 eV$.
  This failure was traced\cite{blmdoom} to the double load borne by
the $\mathbf{\oot}$ FM Higgs irrep  and motivated us to propose
the scenario  that the $\mathbf{10}$-plet and $\mathbf{120}$-plet
shoulder  the main burden of the charged fermion fit, while the
$\mathbf{\oot}$-plet coupling to the matter $\mathbf{16}$-plet  is
small so as to enhance its contribution to the Type I Seesaw (and
lower that to Type II) without any  strong (i.e involving 2-3
generation masses) lower bound on the magnitude of the
$\mathbf{\oot}$-plet coupling due to the requirements of the
charged fermion fit. Although the $\mathbf{120}$-plet has been
considered--somewhat cursorily-- previously in the literature
\cite{early120} and has also recently attracted attention not only
as a perturbation  to correct ``small defects'' in the generic
Babu-Mohapatra(BM) fits\cite{bert} but also as a dominant
contributor to the charged fermion fits\cite{bmsvrad} in
non-supersymmetric theories with radiative neutrino masses
arising\cite{witten}  from exchange of $\mathbf{16_H}$-plet Higgs .
  Yet the combination of $\mathbf{10-120-\oot}$-plets
  with the particular-- and experimentally well motivated --
    assignment of roles,  in the context of
  the renormalizable Supersymmetric $\mathbf{SO(10)}$ GUT,
 that we proposed\cite{blmdoom} has not, to our knowledge,
 been previously  considered  in the literature.

This is the first paper of a series in which we examine
 we examine this scenario\emph{ ab initio} and show that it
enjoys a number of the same virtues manifest in the BM-Type II case
\cite{bsv} and also shown to exist in the
 BM-Type I case\cite{babmacesanu} while being free of
their overall neutrino mass debility. Moreover it leads to a
restrictive type of $\mathbf{b-\tau=s-\mu}$ unification that
radically  reduces the latitude in choosing $m_b(M_X)$ to the
 same level as the uncertainty in $m_{\tau}(M_X)$ yet
 proves not only compatible with $ \mathbf{1\sigma} $ limits but also leads to
 an extremely robust prediction of very near to maximal
  ($\sin^2\theta_{23}^P\geq .95 $) 2-3 generation
 lepton mixing completely compatible with experiment and
 unlikely to be modified when first generation masses and CP
 violation are also accounted for!

We first briefly summarize the essence of the reasons for the
failure of the $\mathbf{10-\oot}$ scenario\cite{gmblm,blmdoom} and
give the fermion mass formulae in the case where a
$\mathbf{120}$-plet is also present.
 We then analyze these formulae in
the case where $\mathbf{10-120}$ plet completely dominate the
charged fermion sector for the \emph{real }two ( 2nd and 3d)
generation case: where a completely explicit and analytic
treatment  is possible. This provides a  insight into the
 the dominant core of the fermion Hierarchy and thus a clear paradigm
 for the analysis of  realistic  3 generation case whose complications
 can be tackled by a perturbative treatment for which
 the $2-3$ sector we analyze is a very stable,robust and non singular
   core and support which dominates the perturbative
   equations up to cubic order in the Fermion hierarchy parameter
   $\epsilon\sim\theta_{12}^c\sim \sqrt\theta_{23}^c \sim  0.2$.
We formulate the fitting equations in a way adapted to determining
their solution by an expansion in    this $\epsilon$ parameter.
 For dominant  2-3 sector neglect of the $\mathbf{\oot}$ couplings
in the charged fermion sector is shown to yield an
acceptable scenario with $\mathbf{b-\tau=s-\mu}$ unification and  maximal
 neutrino mixing and masses for precisely the
 expected values of the 2-3 generation masses at $M_X$
 and a wide and plausible range of the remaining
  parameters ($M_{\nu}^R$ and GUT doublet/MSM doublet Higgs
    fractions) !
  We conclude with an  preview of the  results of  paper II where
the 3 generation real case is analyzed using the perturbation
method based on our results here.

\section{Difficulties of Babu-Mohapatra Seesaw in the MSGUT}

In \cite{blmdoom}  we wrote the Type I and Type II
 seesaw\cite{seesaw} Majorana masses  of the light neutrinos as :

\bea
M_{\nu}^I &=& (1.70 \times 10^{-3} eV) ~ { F_{I}}~
\hat{n}{\sin \beta}\nnu
 M_{\nu}^{II} &=& (1.70 \times 10^{-3} eV) ~{ F_{II}}
 ~\hat{f}{\sin \beta}\nnu
    \hat{n}&=& ({\hat h} -3 {\hat f}) {\hat f}^{-1}
    (  {\hat h} -3 {\hat f})\nonumber
     \eea

where $v=174 GeV$ , $\hat{h},\hat{f}$ are
 the Yukawa coupling matrices of $\mathbf{10,\oot}$ to the
16 plets containing fermion families and  $\beta$ is
 the MSSM Higgs doublet
mixing angle. The functions $F_I,F_{II}$ are defined (upto
irrelevant phases ) as

 \bea F_I &=& {\frac{10^{-\Delta_X}}{2\sqrt{2}}} {\frac{\gamma
g}{\sqrt{\eta\lambda}}} | p_2p_3p_5| {\sqrt{{\frac{z_2}{z_{16}}}}}
\sqrt{{\frac{(1-3x)}{x(1+x^2)}}}
 {\frac{q_3'}{p_5}}\nnu
F_{II} &=& {{10^{-\Delta_X}}} ~{\frac{2\sqrt{2}\gamma
g}{\sqrt{\eta\lambda}}} {\frac{| p_2p_3p_5|}{(x-1)}}
{\sqrt{{\frac{z_2}{z_{16}}}}} \sqrt{{\frac{( x^2 +1)}{x(1-3x )}}}
 {\frac{(4 x-1) q_3^2}{ q_3' q_2 p_5}}\nnu
  \eea

Here $\gamma,\lambda,\eta,g$ are couplings $\sim 1$ while
 the  $p_i,q_i,z_i$ \hfil\break are certain
 polynomials\cite{abmsv,bmsv2,blmdoom}
  in the variable $x $ : which is the  only ``fast'' parameter
  controlling   the GUT scale symmetry
 breaking in the MSGUT\cite{abmsv,bmsv}. This MSGUT is
 based on the $\mathbf{210-10-126-\oot}$ Higgs system
 and is tuned to  keep one pair of Electroweak
 doublets light in the MSSM that emerges
 as the low energy effective theory. The modification(
$\Delta_X= Log_{10} (M_X/GeV) - 16.25 $ ) of
the one loop unification scale due to GUT scale threshold
corrections\cite{ag2} is constrained by the proton lifetime to
be greater than $-1$.

 The `proof' of \cite{gmblm,blmdoom} proceeds by first observing that
in typical BM-Type II fits\cite{gohmoh,bert,babmacesanu} the maximal
value of $ \hat{f}$ eigenvalues is $ \sim 10^{-2}$ while the
corresponding values for $ \hat{h} $ are about $10^2$ times larger.
 As a result  $\hat{n} \sim 10^2 \hat{f}$.
 This implies that  $R={F_I/F_{II}}  \leq 10^{-3}$
  for the pure BM-Type II not
to be overwhelmed by the BM-Type I values it implies. Furthermore
in the BM-Type I fit of \cite{babmacesanu} the requirement of
large neutrino mixing yields, typically, $ \hat{n}_{max} \sim 5
\hat{f}_{max} \sim .3  $ for the maximal eigenvalues of $ \hat{n}$
. Thus $\hat{n}_{max}$ is much smaller than the naive estimate
$(h_{33}- f_{33})^2/f_{33} \sim 10^1$ so that
 magnitudes   $F_I > 10^2$   are required   for
neutrino masses $\sim .05$ eV.  This deserves further
examination on the basis of either a catalog of Type I fits of the
type found in\cite{babmacesanu} or an analytic perturbative
treatment of the fitting problem\cite{nmsgutII,nmsgutIII}. However,
 the reason for the lowering of the eigenvalues of $\hat{n}$ below
  the naive values appears to be the
necessity to have roughly equal elements in the 2-3 sector to
allow large mixing and  so should be robust.

The complete control we have over the
MSGUT\cite{abmsv,ag1,ag2,bmsv,fuku04,fukrebut,bmsv2}  allows us to
demonstrate\cite{gmblm,blmdoom} that both these conditions are not
achievable anywhere over the  complex $x $ plane except where they
also violate some aspect of successful unification (typically the
requirement that $\Delta_X >-1 $ when combined with the above
 requirements of the completely determined coefficient functions
   of the MSGUT ensures the  failure to generate adequately large
   neutrino masses : see \cite{blmdoom} for details).

\section{2 Generation FM fits in  NMSGUTs}

The above  recounted  failure of the Babu-Mohapatra scenario, at
least in generic cases of the MSGUT, could provoke
one\cite{gohmohnasri}
 to entertain more complicated scenarios which give more freedom
  to choose the parameters that enter the
neutrino mass formulae. The likelihood of Type II failure,
already noticed in \cite{gohmohnasri} there motivated the
introduction of additional $\mathbf{54}$ Higgs   as a
work-around of the obstruction posed. However,
 as we already noted in \cite{gmblm},  a proper
demonstration would then require a complete recalculation, of the
various mass matrices and coefficient functions which is --and may
remain-- unavailable, not to speak of the   new
essentially `fast'' parameters introduced by the modified -and
considerably more complex GUT scale  SSB  scenario of such work-arounds.
    To us,  allowing in the arbitrarily
excluded 120-plet, particularly since it does not destroy the
hard-won solution of the GUT scale  SSB problem, is far more
palatable and cogent.

If this is done the  main effect is to introduce two additional
doublet pairs,   from the $(1,2,2) $and $(15,2,2)$  Pati-Salam
submltiplets of the $120$-plet. The GUT scale SSB is
\emph{undisturbed}. The Dirac masses in such GUTs are then
generically given by\cite{ag1,blmdoom}
\bea m^u &=&  v( {\hat h} + {\hat f} + {\hat g} )\nnu
 m_{\nu}&=&v ({\hat h} -3 {\hat f}  + (r_5 -3) {\hat{g}})\equiv
 v ({\hat h} -3 {\hat f}  + r_5' {\hat{g}})
 \nnu
 m^d &=& { v (r_1} {\hat h} + { r_2} {\hat f}  +
r_6 {\hat g}) \\
   m^l &=&{ v( r_1} {\hat h} - 3 {  r_2} {\hat f} +( {{\bar{r}}_5} -
   3\bar{r}_6){\hat g})
       \label{120mdir}\nonumber\eea

 The form\cite{ag1,blmdoom} of the new generic coefficients
$r_5,\bar{r}_{5,6}$ in the particular case of the NMSGUT
is of some interest for future work\cite{csask120}:

\bea \quad  {r_5}&=& \frac{4 i \sqrt{3}{\alpha_5}}{\alpha_6+ i
   \sqrt{3}\alpha_5} \quad
;\quad {{\bar{r}}_5}=  \frac{4 i \sqrt{3}{{\bar{\alpha}}_5}}{\alpha_6+ i
   \sqrt{3}\alpha_5} \cot\beta\quad   \nnu
\bar{r}_6&=& \frac{{{\bar{\alpha}}_6}+ i
\sqrt{3}{{\bar{\alpha}}_5}}{\alpha_6+ i \sqrt{3}\alpha_5}\cot\beta ;\quad
   {\hat g}  = 2ig {\sqrt{\frac{2}{3}}}(\alpha_6 + i\sqt \alpha_5)\sin\beta
 \eea

here $g_{AB}$ is the $\mathbf{120-16-16}$ coupling,and the
 $r_i$ are constants determined  in terms  of the the
 $\alpha_i (\bar{\alpha}_i) $ which are  are the
fractions of the Electroweak doublet $H[1,2,1]$ contributed by the
two [1,2,1]( [1,2,-1]) doublets in the 120-plet .
We have omitted similar details for the `old'  coefficients
  $r_i, \bar{r}_i; i=1,2$\cite{ag2}. Note that with free
$r_i$ our analysis is \emph{generic } and applies to any $SO(10)$
theory with the $\mathbf{10-120-\oot}$ FM Higgs system.

The right handed neutrino mass is
$M_{\bn}  = \hat{f}   \hat{{\bar{\sigma}}}$
and the type I seesaw formula is
\bea M_{\nu}^{I} =   v r_4
\hat{n} \quad ;\quad \hat{n}&=& ({\hat h} -3 {\hat f}) {\hat f}^{-1}
    (  {\hat h} -3 {\hat f})\eea
 where $\hat{{\bar{\sigma}}}     = \frac{i\bar{\sigma}\sqrt{3}}
    {\alpha_2 \sin \beta} $ and $\bar{\sigma} $
is the GUT scale  vev of the $\oot$.

The essence of our proposal is to neglect the presence of $\hat{f}$ in the
Dirac masses  of the 2-3 generations
on the basis of an assumption that $\hat{h},\hat{g}>>\hat{f}$.
 Notice that this implies the strengthening of Type I at the expense of Type II
 Seesaw mechanism.  When one analyzes the complete 3 generation case
  one finds that  dominance of $\hat{h},\hat{g}$  can only be
 strong enough to justify the neglect of $\hat{f}$ to  $O(\epsilon^3)$.
 At  $O(\epsilon^4)$  non zero values of $ \hat{f} $ are
  \emph{required  }in order to preserve the
 consistency of the  charged fermion mass fit at and beyond
 fourth order.  Note that  such   small $\hat{f}$  values  imply
   that the typical  right handed neutrino masses would
   lie in the range  below $ 10^{12} GeV$. Since these masses
   are also thought to obey  the Davidson-Ibarra \cite{davibarra}
   bound  $M_{\nu_R} $ based on  reheating constraints in the
popular leptogenesis  scenario\cite{fukuyana}
 for baryon asymmetry generation
  (which is so natural  in theories supporting   Seesaw neutrino masses).
   This may be a useful clue provided by renormalizable MSGUTs towards
 constraining the parameter space of the leptogenesis scenarios
based on the Type I seesaw mechanism. In any case in the present
2-3 generation Real Core   analysis we will set $\hat{ f} $ to zero
 and take all parameters and Unitary matrices to be real. The
 required generalizations to the 3  generation case are
 quite easy -although computationally tedious - and are
 given in the next paper of this series\cite{nmsgutII}.

The  mass terms above  must be  matched to the
 renormalized mass matrices of the
MSSM evaluated at the GUT scale. While doing so one must
 allow\cite{gmblm}
 for the  possibility  that the fields of the GUT are only unitarily
 related to those  of the MSSM at $M_X$. This introduces several
 $3 \times 3 $
 unitary matrices as well as various phase matrices  into the fitting problem.
 Thus when matching the  charged fermion dirac mass matrices (in which we neglect
 the contribution of the$\mathbf{\oot}$)  we get

\bea m^u &=&  v( {\hat h} +   {\hat g} )   =
  V_u^T  {D}_u Q\nnu
 m^d &=& { v (r_1} {\hat h}    +
r_6 {\hat g})V_d^T  {D}_d R \nnu m^l &=&{ v( r_1} {\hat h} +
r_7{\hat g})= V_l^T  {D}_l L\nnu
 r_7 &=& \bar{r}_5 - 3 \bar{r}_6  \eea

Where $D_{u,d,l}$ are the charged fermion masses at $M_X$  and
$V_u,Q,V_d,R=C^{\dagger} Q,L,V_l$ are arbitrary unitary matrices
($C$ is the CKM matrix) which are to be fixed by convention
--where allowed by conventional ambiguities -- or determined by
the fitting procedure in terms of the low energy data, or left as
parameters to be determined by future experiments sensitive to
degrees of freedom and couplings (e.g baryon violating couplings)
that the low energy data is not.

To put our equations in a form transparent enough to clearly
separate the contributions of the $\mathbf{10,120 }$ (and
-eventually-  $\mathbf{\oot}$) plets\cite{nmsgutII} we write the
matrices $V_{u,d,l}$ associated with the anti-fermion fields  as
new unitary matrices $\Phi_{u,d,l}$ times the fermion field
matrices $Q,R,L$. \bea V_d &=& \Phi_d R \hspace{5mm}; \hspace{5mm}
V_u = \Phi_u  Q \hspace{5mm}; \hspace{5mm} V_l = \Phi_l L \eea We
then separate symmetric and antisymmetric parts for each charged
fermion equation
\bea Z &=& \Phi^T {D} + {D} \Phi \hspace{5mm};
\hspace{5mm} A = \Phi^T D - D \Phi  \eea

Then it is easy to solve for the matrices $\hat{h},\hat{g}$ using
(say) just the d-type quark equations and substitute in the other
two equations
 to obtain the equations in the   form
\bea
C^* Z_d{C}^{ \dagger}&=& r_1 Z_u\hspace{5mm} ;
 \hspace{5mm} Z_d = D Z_l D^T \hspace{5mm};\hspace{5mm}
D = R^* L^T  \\
C^* A_d{C}^{ \dagger}&=& r_6 A_u \hspace{5mm} ;
\hspace{5mm}r_7 A_d = r_6 D A_l D^T \label{aseq}\eea

In the 2 generation case the  antisymmetric  equations(\ref{aseq}) serve
only to fix the parameters $r_6,r_7$ and thus play no further role.
On the other hand since $C,D$  are orthogonal matrices it follows that
$r_1=Tr(Z_d)/Tr(Z_u)$   so that we can write these two equations
 in the (dimensionless)  form

\bea
\widehat{S}_1 &=& \frac{C Z_d C^T}{Tr Z_d}  - \frac{  Z_u  }{Tr Z_u}=0\nonumber\\
\widehat{S}_2 &=& \frac{Z_d - D Z_l D^T}{Tr Z_d}=0
\eea

The convenience of this form is that the hierarchy within each generation
will now structure the problem and render its solution almost trivial in the
two  generation case and tractable in the 3 generation case. It is amenable
 to an expansion in the hierarchy parameter $\epsilon$ which reduces the problem
 of finding  a sufficiently accurate fit to the $\sim 10^4 $ intra-generational
 variation in mass to an exercise in going to sufficiently high order in
 the expansion parameter $\epsilon$\cite{nmsgutII}.

\subsection{Analytic Solution of the Charged Fermion Mass Relations }

We parametrize the matrices  $\Phi_{u,d,l},C, {\cal{D}}$ as \bea
\Phi_{u,d,l} &=& \left( \begin{array}{cc}\cos\chi_{u,d,l}
\hspace{5mm}\sin\chi_{u,d,l}\\
-\sin\chi_{u,d,l} \hspace{5mm}\cos\chi_{u,d,l} \end{array} \right) \nonumber\\
 {C,\cal{ D}} &=& \left( \begin{array}{cc}\cos\chi_{c,\cal{D}}
\hspace{5mm}\sin\chi_{c,\cal{D}}\\
-\sin\chi_{c,\cal{D}} \hspace{5mm}\cos\chi_{c,\cal{D}} \end{array}
\right)  \eea
 and define   $\Delta u=u_3-u_2~,~ T_u=u_3 + u_2 ~,~ \Delta
d=d_3-d_2~,~ T_d=d_3 + d_2~,~\Delta l=l_3-l_2~,~ T_l=l_3 + l_2$.

 Then the 2 independent  equations  in $\hat{S}_1=0$
reduce to \bea \tan \chi_u &= & \tan (\chi_d - 2 \chi_c)\nnu
\cos\chi_d &=& \alpha \cos (\chi_d - 2 \chi_c) \\ \alpha &=&
{\frac {(\Delta d ) (T_u)}{ (T_d) (\Delta u)}}={\frac {(d_3-d_2 )
(u_3+u_2)}{ (d_3+d_2) (d_3-u_2)}} \nonumber\label{chiud}\eea
Where $\chi_c$ is the  Cabbibo angle in the 2-3 sector i.e
 in the toy CKM matrix $C$ and $u_{2,3},d_{2,3},l_{2,3}$ are
the charged fermion  masses of the 2nd and 3d generations \emph{
up to arbitrary signs}. Since $\chi_u =   \chi_d - 2 \chi_c$,
 and $\chi_c\sim O(\epsilon^2) $, $ \quad \chi_{ud}\approx\chi_{d}$
 to   order in $\epsilon^2$.
The second equation in (\ref{chiud}) equation can be solved for
$\tan\chi_d$ : \bea \tan \chi_d &=& \frac{\csc 2 \chi_c}{\alpha}-
\cot 2\chi_c \label{chid}\eea Since $\alpha=1+(\epsilon^2)$, it is
clear that the leading ($\sim \epsilon^{-2} $) contributions
cancel leaving behind an $O(1)$ result.
 Similarly the \emph{three}  equations $\hat{S}_2=0$ yield
\bea \tan (\chi_l - 2 \chi_D) &= & \tan \chi_d\nnu \cos\chi_l
  ={\frac{T_d}{T_l}}  \cos  \chi_d  &=&{\frac{d_3 +d_2}{l_3+l_2}}
\cos \chi_d   \nnu \label{chidchil}\eea
Thus
 \bea\chi_D&=&(\chi_l-\chi_d)/2 \qquad;\qquad
\chi_l=\pm\bar{\chi}_l \nnu \bar{\chi}_l&=&\cos^{-1} {\frac{T_d
\cos \chi_d}{T_l}}\eea
To leading order $\chi_l =\pm \chi_d$. However note at at next to
leading order the ratio $T_d/T_l$ must not become so large
($\tan\chi_d^0\sim .9$) as to make $\chi_l$ the inverse cosine of
a number bigger than 1.

 The third equation yields the all important
consistency condition
 \bea { {d_3 - d_2  =\Delta d = \Delta l  = l_3 -l_2 }} \eea
i.e
\bea  \mathbf{{m_b(M_X)  - m_{\tau}(M_X)  = m_s(M_X)  -
m_{\mu}(M_X) }}\eea
This is one of the main results of this paper. It should be
understood that there is a sign/phase ambiguity--arising from the
arbitrariness in the unknown Unitary matrices that     relate the
SGUT and MSSM fields (see above)-- for each of the masses, while
their magnitudes are determined  by the RG flow of the MSSM into
the UV.

\section{$\mathbf{b-\tau=s-\mu}$ Unification}

Several questions arise regarding the status of this
version of $b-\tau $ unification which is imposed
by the consistency of the assumption that the $\mathbf{10,120}$
 multiplets dominate the heavy fermion mass matrices.
  Firstly it should be noted that the necessity   to the zeroth
order(in $\eta=\epsilon^2$) version of this relation
i.e $m_b(M_X)  - m_{\tau}(M_X)=0$ was already noticed by
Bajc and Senjanovic\cite{bmsvrad}   while studying a technically related
model but with a   but quite different (\emph{non-supersymmetric})
scenario.   In that model the $\mathbf{10-120 }$ are used to fit
the 2-3 generation  masses and   neutrino masses are generated
\emph{radiatively }\cite{witten} due to    exchanges involving a
a SO(10)  \emph{spinorial Higgs} $\mathbf{16}_H$-plet. Thus the
motivation, origin and properties of that model are
totally distinct from our   proposal which is supersymmetric
(and thus protected from radiative corrections of its
Superpotential), employs no  Higgs  16-plet and  has a tree level
Type I seesaw using the $\mathbf{\oot}$ (which also \emph{necessarily}
contributes to the fermion masses at order $\epsilon^4$\cite{nmsgutII})
in the three(\emph{but not two})  generation case.
We have here analyzed completely   the two generation case and
to order $\epsilon^6 $ for the 3  generation (CP conserving case )
in \cite{nmsgutII}.

      The first question that arises is whether this relation will
be preserved when the first generation and then CKM  CP violation
are also introduced.
The explicit analytic solution trivially derivable from the
 above equations  given above coincides, order by order
in perturbation theory , \emph{including the
$\mathbf{b-\tau=s-\mu}$ constraint }, with the solution found
\cite{nmsgutII}  by expanding all angles $\chi$ in powers of
$\epsilon$  and solving the equations $\hat{S}_1=0=\hat{S}_2 $
 order by order in $\epsilon $ (and truncating all expansions by dropping
 undetermined coefficients left after solving to desired order). The
 $\mathbf{b-\tau=s-\mu}$ unification constraint arises at order
 $\eta=\epsilon^2$.  In the three generation case the analytic solution is not available but
 the complete regularity observed in the perturbation theory in
 $\epsilon $  and the extreme smallness of the first
generation perturbations  makes an expansion in $\epsilon$ well motivated.
 Then we find that the  $b-\tau=s-\mu$ unification constraint is
 reproduced at $O(\epsilon^2) $.
One may also expect that  this constraint will persist when the CP
violation is introduced and a perturbation theory in $\epsilon$
used to solve the the fermion mass relations   since it is well
known that CP violating effects are Wolfenstein suppressed. In
that case, however, the relation
 will also contain phase uncertainties generalizing the sign
 uncertainties  present in the real case.
\emph{Thus $\mathbf{b-\tau=s-\mu}$ unification may be considered
as one of the  signature ``tunes'' of the fermion mass (FM) fit in
the NMSGUT  scenario }we proposed\cite{blmdoom} and have begun
elaborating on here.

Moreover the pursuit of the perturbation theory to $O(\epsilon^3)$
 gives\cite{nmsgutII} an additional remarkable constraint :
 \bea \mathbf{\theta^c_{13} = \theta^c_{12}\  \theta^c_{23}} \eea
  between the  CKM rotation angles !
   No further parameter constraint was found
 up to 5th order in $\epsilon$. However at 4th order we
 found that the contribution  of the $\mathbf{\oot}$-plet becomes
  \emph{necessary }to avoid inconsistency
 in the $ \hat{S}_1 , \hat{S}_2 $ fitting equations, which
 remain combined in a  single equation
 $ a  \hat{S}_1 +  b  \hat{S}_2 =0 $
 rather than  vanishing separately, while the coupling of the
 $\mathbf{\oot}$-plet is determined to be
 $\sim  \hat{S}_2 \sim \epsilon^4$.
 Given enough computer power we see no reason
 why the perturbation cannot be carried to sixth or even seventh order in
 $\epsilon$ : which should be more than sufficient to ensure convergence
 to the  accuracy to which the lepton masses at $M_X$ are known :
 given our   uncertainty regarding
 the crucial quantities $M_S, \tan\beta$.

Since $d_2/l_3,l_2/l_3,\chi_c \sim\epsilon^2$ and
$u_2/u_3\sim\epsilon^4$ it immediately follows that the  solution
of eqn(\ref{chiud}-\ref{chid}) to leading order in epsilon is
\bea \tan\chi^0_u =\tan \chi^0_d ={\frac{d_2}{l_3 \chi_c}}
+O(\epsilon^2) \eea
Thus the magnitude  of $\tan \chi^0_d$ is determined to lie (see
data values below) in the range
\bea |\tan \chi^0_d|  = .4974 (+.19) (-.1) \eea
 Thus  the accuracy
of the leading order  is $\sim 10\%$, while the experimental
uncertainties are larger. For all practical purposes we may
eliminate $l_3$ in favour of $ \tan \chi^0_d $  as the  basic
parameter and let it vary in the range above  with either positive
or negative sign. Note that these values will prove  crucial in
showing that the leptonic mixing is near maximal over a wide range
of parameters.

\subsection{Numerical Values}

 Even the very precisely known low energy
 lepton masses become smudged by uncertainties when RG evolved past the
 unknown thresholds  of order $M_S$ and   $M_{\bar{\nu}}$
 with unknown $\tan\beta$. Since the
 effects of CP violation on the $O(\epsilon^3)$
  relation $\mathbf{\theta^c_{13} = \theta^c_{12} \theta^c_{23}}$
   may  well
  be non-negligible,  the precise numerical implementation of the
 constraints seems premature. Still a  survey of the typical
 fermion mass-mixing data sets is quite comforting since it
 exhibits  at least approximate compatibility with the constraints
 found by us and makes clear the huge parameter space available
 for satisfying them. Moreover implementation of
 the necessary $\mathbf{b-\tau=s-\mu}$ constraint implies a radical
  ($10-100 $ fold !)  reduction in the uncertainty
   allowed in $m_b(M_X)$ due to its
 strict correlation with $m_{\tau}(M_X)$ : which is
 much more precisely known(see below).

In \cite{dasparida} the following values are given for the two
loop renormalized running fermion masses (in units of GeV) at
$\mu=M_X=2 \times 10^{16} GeV $ for the representative high  $\tan
\beta$ value $\tan \beta (M_S=1TeV) =55 $
\bea
 m_u  &=& (0.7244^{+0.1219}_{-0.1466})\times 10^{-3}\ ;
 \ m_c = .2105049^{+.0151077}_{-.0211538}\ ;\quad\quad
 \ m_t  = 95.1486^{+69.2836}_{-20.659} \nnu
 m_d  &=& (1.4967^{+0.4157}_{0.2278}\  )\times 10^{-3} ;
  \ m_s  = (29.8135^{+4.1795}_{-4.4967} )\times 10^{-3} ;
   \ m_b  = 1.4167^{+0.4803}_{-0.1944} \nnu
   m_e&=&( 0.3565^{-0.001}_{+0.0002})\times 10^{-3}  ;
   \ m_\mu  = (75.2938^{-0.1912}_{+0.0515})\times10^{-3}\  ;
    \ m_\tau  = 1.6292^{+0.0443}_{-0.0294} \nnu
     \tan\beta&=& 52.0738^{-16.5475}_{+4.3757}\  ;\quad\quad\quad
      \ v_u  =117.7947^{-46.7214}_{+19.2752}\  ;\quad\quad\quad
        \ v_d   = 2.2620^{-0.2615}_{+0.1661} \nonumber \eea
It is apparent that after  including the sign ambiguity $|d_2-l_2|$ can
be anywhere from $.046 GeV$ to $.105 GeV$ with an notional
accuracy  limited by the low energy error induced  uncertainty in
$m_s(M_X)$  of about $.004 GeV$. On the other hand $|d_3-l_3|$  is
about $.2$  GeV$ \pm .5$ GeV (!) (evidently $d_3,l_3$ must have
the same sign !).  There is a long way to go before the validity
of the constraint can be verified, even granted that no additional
ambiguities due to CP violating phases arise.

Our approach is thus to take  as reference central values the
relatively  accurate values of $m_{\mu,\tau,s}$, at large
$tan\beta$ (obviously favoured by SO(10) GUTs) (say $\tan \beta
=55$ for definiteness) and assume $d_3=m_{\tau} \pm m_s \pm
m_{\mu}\approx 1.63  +sign[d_2] .03  - sign[l_2] .075) $ to
satisfy the consistency constraint,
  secure in in the knowledge that the data allows $m_b$ to
lie in the range $(1.23 , 1.9)$ for $\tan\beta =55$ and somewhat
lower values for lower $\tan \beta$. However we find that
$sign[d_2]=+ , Sign[l_2]=- $ is impermissible since it leads to an
imaginary value for $\chi_l$ when one solves eqn(\ref{chidchil}).

   The values of $d_3$
required in our  scenario by the combination of $b-\tau=s-\mu$
unification and near maximal  leptonic mixing tend (at $d_3\sim
1.585$) to be somewhat larger than the central value quoted for
$m_b$ above but still comfortably within the $ \mathbf{1\sigma}$ error
bars. This can  be considered a prediction of this model which may
become constraining as limits improve. Note particularly that the
implication of this type of  novel constrained unification
leaves only the relatively tiny $ \mathbf{1\sigma}$
 uncertainty ($\sim .001  - .05 GeV$ )  in $m_{\tau}(M_X)$
 as a fudge factor while $d_3$
 whose  (i.e $m_b(M_X)$'s)  uncertainty  could have
  been 10-100 times larger is tied down to vary in
  tandem with $m_\tau(M_X)$. Yet we will see that the prediction
   of the maximal mixing   is still extremely robust.

 The values of the mixing
angles at $M_X$ given by \cite{dasparida} (who assume
$\delta_{CKM} =\pi/2$   for convenience ) are $\theta_{12}=.221
,\theta_{23}= .037,\theta_{13}=.003$, with no uncertainty quoted.
The product $\theta_{12} \theta_{23}= .008\pm .002$ is of the same
order of magnitude as $ \theta_{13}$. Since this relation arises
in the 3 generation case at $O(\epsilon^3)$, by which stage the CP
violation  also enters it seems legitimate to hope for a
considerable ($\sim 100\%$) role for $\delta_{CKM} $ in modifying
the relation $\mathbf{\theta^c_{13} = \theta^c_{12}
\theta^c_{23}}$ between angles.

\section{Neutrino Masses and Mixing}

In this section we will first determine the   angle in the
 PMNS\cite{pontemaki} mixing matrix and then proceed to
 show that this mixing angle can can easily satisfy the
 near maximality constraints\cite{strumviss} imposed by experiment.

\subsection{PMNS Matrix}
Under the  assumption that the couplings $\hat{h},\hat{g}$ of the
$\mathbf{10}$ and $\mathbf{120}$  completely dominate coupling  $\hat{f}$  of the
$\mathbf{\oot}$,  the Type I seesaw masses in this theory are given by

\bea M_{\nu}^I&\simeq & -\frac{v^2}{2\widehat{\bar{\sigma}} }
(\widehat{h}+
r_5^{'}\widehat{g})^T \hat{f}^{-1}(\widehat{h}+ r_5{'}
\widehat{g})\nonumber\\
 &=& r_4' R^T \left( \frac{Z_d}{ r_1}+ \frac{r_5^{'}}{
r_6}A_d\right)^T R S^{-1} D_{\hat{f}}^{-1}S^{-1T}R^T
\left(\frac{Z_d}{  r_1}+ \frac{r_5^{'}}{  r_6}A_d\right)R
\nonumber\\
&\equiv&  r_4  R^T Y_d^T R S^{-1}
 Diag(1,\rho) S^{-1T}R^T Y_d R\nnu
 &\equiv& R^T F R
\nonumber\\
&=&  {L}^T \mathcal{P} D_{\nu}\mathcal{P}^T  {L} \label{TypeI}\eea
Where ${\cal{P}}$ is the Lepton mixing (PMNS) matrix in the basis
with diagonal leptonic charged current, $D_{\nu}$ the light
neutrino masses extrapolated to $M_X$, and $Y_d$ a
convenient dimensionless form of the
linear combination of $Z_d,A_d$ that determines the PMNS mixing.

  The explicit form of $Y_d$ is
\bea Y_d  =\left( \begin{array}{cc} {\frac{2 d_2 }{\Delta
d}}\cos\chi_d
  \hspace{10mm}
   - \sin\chi_d - r_5'{\frac{T_u }{\Delta u}}\sin\chi_u  \\ \\
 - \sin\chi_d + r_5'{\frac{T_u }{\Delta u}}\sin\chi_u
 \hspace{10mm} 2 (1 + {\frac{ d_2 }{\Delta d}}) \cos\chi_d
 \label{pmns}\end{array}\right)\eea
So far we have not used the family basis ambiguity of the SO(10)
GUT  that allows us to perform an arbitrary unitary redefinition
of the matter $\mathbf{16}$-plets at will. Since $\hat{f}$ is
symmetric it may be written as $\hat{f} =S^T D_{\hat{f}} S $ where
S is unitary and $D_f$ is diagonal and real. The basis ambiguity
can be fixed by a choice of S to be any given unitary matrix : for
example to be unity. Here we make the convenient choice $S=R$
since this removes the obscuring factors $RS^{-1}$ in the Type I
mass formula eqn(\ref{TypeI}) above, leaving behind the two
eigenvalues of $f$ as free parameters. One of these can be
extracted into the overall scale say $ \hat{f}_2 $ leaving behind
the parameter $\rho=\hat{f}_2/\hat{f}_3 = M^{(2)}_{\bar{\nu}}/
M^{(3)}_{\bar{\nu}}$ to affect the determination of the mixing
angle $\chi_{\cal{P}}$.

Imposing the family  basis choice and diagonalizing the
matrix $F$ we immediately obtain the neutrino
mixing and masses  to be
\bea
 F&\equiv& Y_d^T Diag(1,\rho) Y_d  \equiv  {\cal{F}}
Diag(F_2 , F_3) {\cal{F}}^T \nnu
{\cal{P}}& =& D^{\dagger} {\cal{F}} \qquad ;\qquad D_{\nu}
={\frac{r_4}{\hat{f}_2}} Diag(F_2 , F_3) \eea
 The mixing angles $\chi^{\pm}_F$ of the 2-d rotation matrix
 ${\cal{F}}$  which diagonalizes $F$ to\hfil\break
  $ Diag (F_{\mp},F_{\pm})$  are  given  by
 \bea
 \chi_F^{\pm} &=& \tan^{-1} (  \omega_F \mp {\sqrt {1 +
  \omega_F^2}} )\nnu
\omega_F&=& {\frac{F_{11}-F_{22}}{ 2 F_{12}}}\eea
   The eigenvalues $F_{\pm} $obey
 \bea
 F_{\pm} &=& {\frac{1}{2}}( Tr(F) \pm {\sqrt{tr(F)^2
-Det[F]}})\nnu
sign[Tr(F)] &=&   sign[F_{+}^2-F_{-}^2)] \eea
 Note that since  (to leading order in $\epsilon$)
 \bea Tr(F)= \sin^2\chi_d(4
+ 16 \rho +r_5^2(1+\rho) -4 r_5(1+2\rho) +
 4\rho \cot^2\chi_d )\eea
 the hierarchy will invert for
 \bea
 \rho < -{\frac{(r_5-2)^2}{(r_5-4)^2 + 4 \cot[\chi_d]^2}}\equiv\rho_{inv}
 \eea

\subsection{Maximal 2-3 PMNS Mixing}
We now have a formula for the leptonic mixing in terms of two
quite arbitrary parameters $\rho, r_5=r_5'+3 $.
  The value  $\rho=  1$ corresponds
   to completely degenerate right handed (heavy)
neutrinos.   While $|\rho| <<1 $ and $|\rho|>>1 $ correspond to
hierarchical righthanded neutrino masses,  $r_5=0$ corresponds to
negligible contribution from the SU(4) singlet pair of MSSM
doublets from the $\mathbf{120}$-plet (i.e Pati-Salam
sub-representation representation $(1,2,2)\subset \mathbf{120}$)
to the low energy symmetry breaking. Thus it may be  realizing  a
Georgi-Jarlskog mechanism\cite{georjarls} for the
$\mathbf{10\oplus 120}$  based fit of the 2-3 generation charged
fermion masses.    We will show that

\begin{itemize}

\item For \emph{any}  value of $r_5$  there exists a range of values
of $\rho$ for which the  the leptonic mixing is  near maximal
($1\geq \sin^2 2 \chi_P\geq .9$).

\item There exist \emph{ranges} of $r_5$ where the width of the
 $\rho$-band where  $1\geq \sin^2 2 \chi_P\geq .9$  becomes very large.
\end{itemize}
 Thus large mixing is \emph{always} achievable and
 \emph{inevitable} in certain broad regions of the $r_5,\rho $ plane.

 From  the analytic solution of the charged fermion
 fit (eqns(\ref{chiud},\ref{chidchil}) and eqn(\ref{pmns}) we find that
 the two possible values of the total mixing are

\bea\chi^{(0)}_P=\chi^-_F  \qquad\qquad;\qquad\qquad
\chi^{(1)}_P=\chi^-_F
 + \chi_d \eea
So that the  two possible values of the mixing parameter
 $\sin^2 2 \chi_P $ (which is indifferent to inversion of the
 $\nu_L$ hierarchy ) are
 \bea
\sin^2 2 \chi^{(0)}_P &=& {\frac{1}{(1+\omega_F^2)} }\nnu
 \sin^2 2 \chi^{(1)}_P &=& {\frac{ \left(-1 + 2\omega_F \tan\chi_d  +
   \tan^2\chi_d\right)^2} {\left(1 +\omega_F^2 \right)
   \left(1 +\tan^2\chi_d \right)^2}}\label{ssq2theta}\eea
$\sin^2 2 \chi^{(0)}$ is maximal at $\omega_F=\mathbf{\omega^{(0)}_F=0}$,
 while $\sin^2 2 \chi^{(1)}$ is maximal at
\bea\omega_F= \omega^{(1)}_F =
{\frac{2 \tan\chi_d}{2 \tan^2\chi_d -1 }} \eea
Moreover the range of $\omega_F$  values which leads to
 near maximal mixing ($1\geq \sin^2 2 \chi^{0}_P\geq .9$) is
 $ \omega_F\in (-1/3,1/3)$. While for
 $1\geq \sin^2 2 \chi^{1}_P\geq .9$ this range  lies between
\bea \omega^{(1,\pm ,.9)} = {\frac{ \pm t^2  + 6\,t \mp 1 } {
3\,t^2  \mp 2\,t -3}}\eea
 There are in principle  different combinations of signs :
$sgn\{d_2,l_2,u_3,u_2,\chi_c\}$   that can be chosen  for the various
 parameters. They make no difference to the mixing
at the leading order. However,   the combination
$d_2$ positive, $ l_2$ negative,
leads to \emph{imaginary} $\chi_l $ in the full theory(i,e beyond leading
order where $\chi_l=\pm\chi_d$)  and hence must
be discarded. The  remaining sign choices give results that are
only marginally different from those where we take all masses and
$\chi_c$ as positive. So will not here  discuss their minor differences
and focus only on the ``all positive'' sign choice in the interests of clarity.

\subsection{Depiction of the Large Mixing Parameter Regions}

The regions of the $r_5,\rho$ plane that support large PMNS mixing
can be fairly completely and accurately delineated
 by working to  leading order in $\epsilon$. Then
 $\tan\chi_d\approx m_s/\chi_c m_\tau$  \emph{is the only input
 parameter required}. In leading (and determinative)
  order   the formulae  are  not obscured by the other
    parameters whose role in determining
 the PMNS mixing is quite marginal. Note, however, that
 since we have an exact analytic solution numerical work can
 always use  it to confirm  and refine the  features
 visible   at leading order.

It is easy to show that to leading order in $\epsilon$
\bea \omega_F = \frac{\left( -{\left( -2 + {r_5} \right) }^2 +
 {\left(-4 +  {r_5} \right) }^2\,\rho -
      4\,\rho\,{\cot (\chi_d)}^2 \right) \,\tan (\chi_d)}{4\,
      \left( -4 + {r_5} \right) \,\rho}\eea
Note particularly the singularities at $\rho=0$ and at $r_5=4$ and
the fact that the result is dependent only on $r_5,\rho,\tan
\chi_d$.  There is clearly never a large mixing $\chi_P^{(0)}$
solution for $\rho=0$ or $r_5=4$, and it is easy to check that for
these values the limit as $\omega\rightarrow\infty,
  \sin^2 2 \chi^{(1)}_P \rightarrow
  \frac{4\,\tan^2\chi_d}{{\left( 1 + \tan^2\chi_d\right)}^2}$
    which never yields large mixing for $ \tan\chi_d \in (.35,.6)$
    which is the experimentally allowed range.
Clearly eqn(\ref{omegaF})  implies that for arbitrary
$(r_5,\tan\chi_d)$ one can\emph{ always} find a value of
$\rho_{max}^{(0)}$ where the mixing $\sin^2 2 \chi^{(0)}_P $    is
maximal. :
\bea \rho_{max}^{(0)} ={ {\frac{{\left( -2 + {r5} \right) }^2}
{{\left( -4 + {r5} \right) }^2 - 4\,{\cot
({\chi_d})}^2}}}\label{omegaF}\eea
 For example, when $r_5=0, \tan\chi_d=.4974$ ,
$\rho_{max}^{(0)}=-23.8511$ (while $\rho_{inv}=-0.124348$, so the
hierarchy  is  inverted but not acutely so). Moreover even for
$\rho=1,r_5=0$ (degenerate $\nu_R$, Georgi-Jarlskog point) one
finds \bea \omega_F={\frac{1}{4}} (\cot\chi_d - 3 \tan\chi_d) \eea
which gives $\sin^2 2\chi^{(0)}_P =.9835$  at $ \tan\chi_d= {
{d_2}/{(l_3 \chi_c)} }= .4974 $. Giving already a hint of the
robustness of large mixing.

 For given $r_5,\tan\chi_d$ the edges $\rho^{(0,\pm 1/3)}$
of the large mixing band for the mixing $\chi^{(0)}_P$  are
\bea \rho^{(0, +,.9)} &=& \frac{3\,{\left( -2 + {r5}
\right) }^2\,\tan ({\chi_d})}
  {-4\,\left( -4 + {r5} + 3\,\cot ({\chi_d}) \right)  +
    3\,{\left( -4 + {r5} \right) }^2\,\tan ({\chi_d})}\nnu
\rho^{(0,- ,.9)} &=& \frac{3\,{\left( -2 + {r5} \right) }^2\,
\tan ({\chi_d})}
  {-12\,\cot ({\chi_d}) + \left( -4 + {r5} \right) \,
     \left( 4 + 3\,\left( -4 + {r5} \right) \,\tan ({\chi_d}) \right) }
     \eea
while the width of the band is
\bea
\Delta^{(0,.9 )}&=&
 \rho^{(0, {\frac{1}{3}})} - \rho^{(0,  - {\frac{1}{3}})}\nnu
&=&\frac{24\,\left( -4 + {r5} \right) \,
{\left( -2 + {r5} \right) }^2\,\tan ({\chi_d})}
  {144\,{\cot ({\chi_d})}^2 + {\left( -4 + {r5} \right) }^2\,
     \left( -88 + 9\,{\left( -4 + {r5} \right) }^2\,
     {\tan ({\chi_d})}^2 \right) }\eea
An exactly parallel discussion can be given for the other
case, we omit the tedious details.

The (simple) poles  of $\Delta^{(0,.9 )}$ are at $r5 \rightarrow
-1.57871, 1.10189, 6.89811, 9.57871 $. In Fig.1 we plot the width
of the large mixing stripe versus $r_5$ to illustrate its large
variation. It is apparent that it is precisely the poles of
$\Delta^{(0,.9 )}$ that cause the  width to become large and that
the whole region between $r_5=-1.6 $ and $r_5 = 10 $(particularly
$r_5\in (6.5,10)$ supports very large regions where the mixing is
maximal.

\begin{figure}[h!]
\begin{center}
\epsfxsize 15cm \epsffile{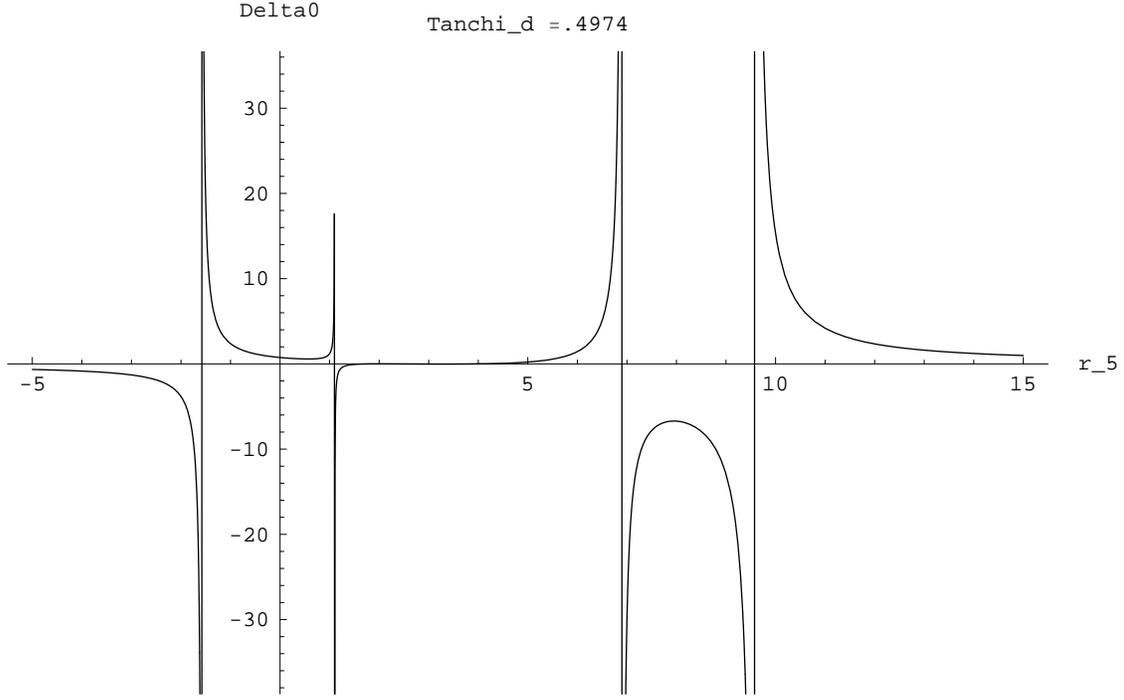}
 \caption{  Plot  of  $\Delta^{(0,.9 )}$ vs $r_5$,  at lowest
 order in $\epsilon$,    at fixed  $\tan\chi_d =.4974$.
  The poles of $\Delta^{(0,.9 )}$ cause regions of a very wide
   range of $\rho$ to support large PMNS mixing.}
\end{center}
\end{figure}

This is clearly seen in the corresponding contour plot Fig. 2
 where we have truncated the extent in the $\rho$ direction to avoid
obscuring the small mixing band around  $\rho=0$.

\begin{figure}[h!]
\begin{center}
\epsfxsize 15cm \epsffile{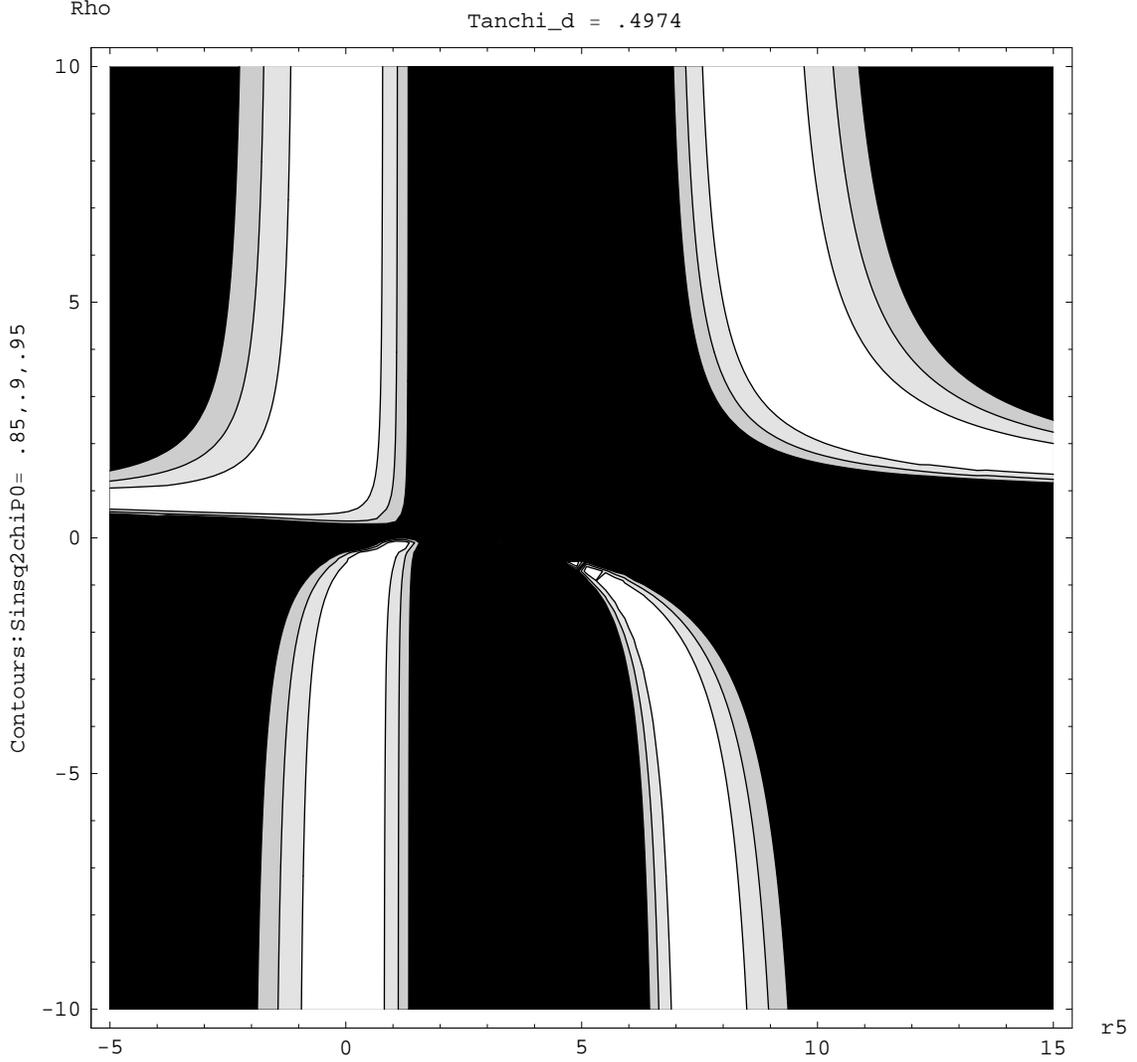}
 \caption{Contour Plot  of the mixing parameter
 $Sin^2 2 \theta_{23} =Sin^2 2  \chi_P^{(0)}$, to lowest
 order in $\epsilon$,  on the $(r_5,\rho)$ plane at fixed
 $\tan\chi_d$. The contours shown are at
 $Sin^2 2 \chi^{(0)}_P=.85,.9,.95$ with the shading
 becoming darker towards smaller values. }
\end{center}
\end{figure}

An exactly parallel discussion can be given for the mixing $ Sin^2
2  \chi_P^{(1)}$, we omit the details and give only the contour
plot in Fig. 3, which has an analogous structure to that for
$Sin^2 2 \chi^{(0)}_P$. Note that the the very narrow width of the
large mixing band for $r_5\in(1.3,5)$ has caused it to become
invisible due to the large range on the $\rho$ axis. It is easy to
see this relatively narrow band by magnifying the plot. In Figure
4  we show an example of the variation of the mixing parameter
with $\tan\chi_d$ keeping $r_5,\rho$ fixed at values in the
maximal mixing regions ($\sin^2 2 \chi^{(0 )}_P \geq .95$).

\begin{figure}[h!]
\begin{center}
\epsfxsize15cm\epsffile{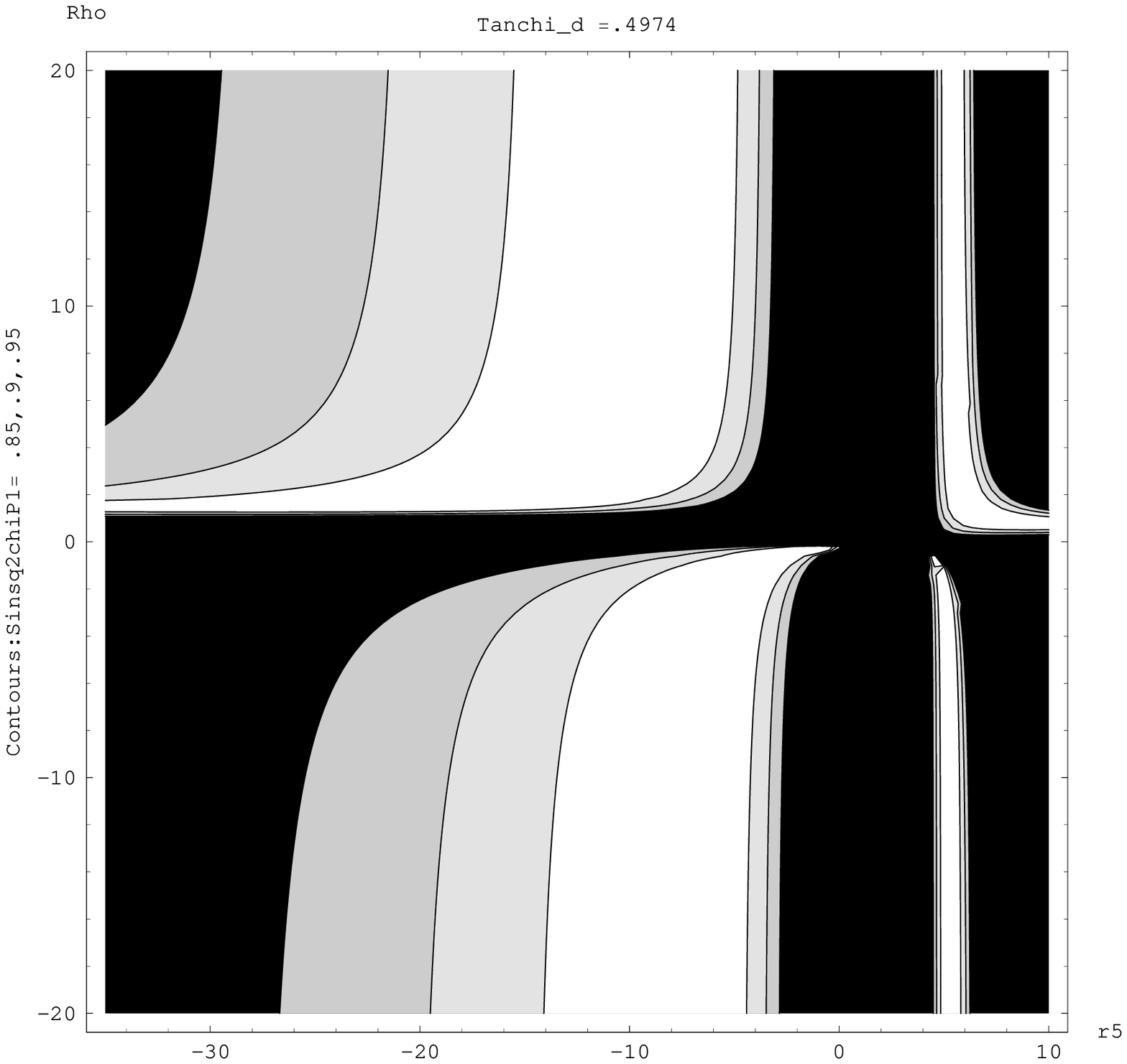}
 \caption{ Contour Plot  of the mixing parameter
 $Sin^2 2 \theta_{23} =Sin^2 2  \chi_P^{(1)}$, to lowest
 order in $\epsilon$,  on the $(r_5,\rho)$ plane at fixed
 $\tan\chi_d$. The contours shown are at
 $Sin^2 2 \chi^{(0)}_P=.85,.9,.95$ with the shading
 becoming darker towards smaller values.  }
\end{center}
\end{figure}
\begin{figure}[h!]
\begin{center}
\epsfxsize15cm\epsffile{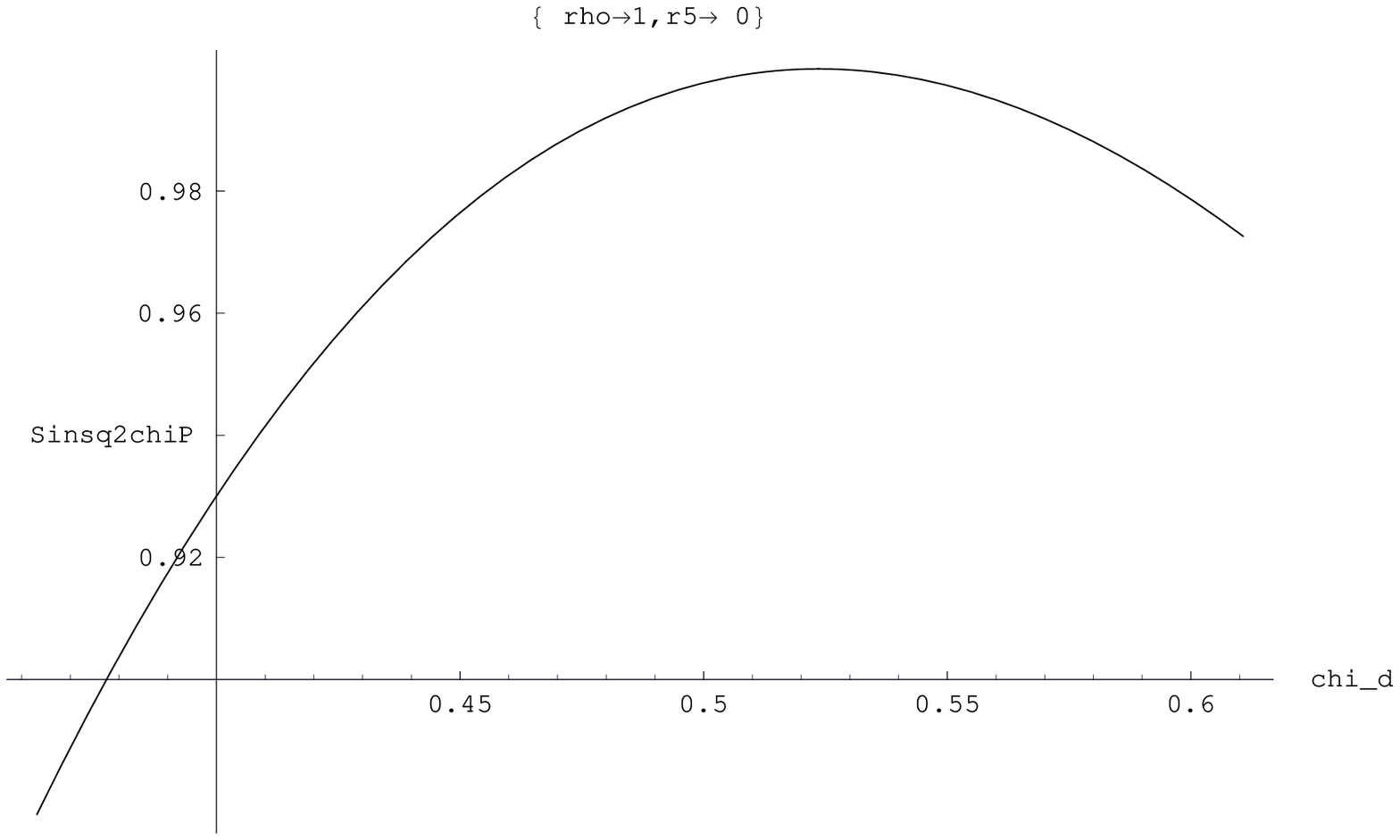}
 \caption{ Plot of the mixing parameter
 $Sin^2 2 \theta_{23} =Sin^2 2 \chi_P^{(0)}$
 , to leading order  in $\epsilon$, versus $ \chi_d$ at fixed
 $r_5,\rho$.}
\end{center}
\end{figure}
  Although the exact formulae given above have a completely
  regular expansion in the small parameter $\eta=.04$  it
  is of interest to see how much the leading order results
 are modified when we include all orders in $\eta$ and
  also to examine the effects of varying the other
 charged fermion masses  e.g $ m_{top} $.
 The explicit formulae -although trivial to write down-
 are not very transparent . However   plots
(Figs 5-7) analogous to the ones for the leading order formulae
tell almost the the whole story :

\begin{itemize}

\item  The contour plots for the exact results are essentially
   identical to those for the leading order case apart from minor
    changes in fine structure and a slight squeezing of the maximal
    mixing regions. The similarity is so great that we show only
    one example as Fig. 5 : the contour plot
    $Sin^2 2 \chi_P^{(1)}$. We  show  the plot  for a
    different sign choice  for $d_2,l_2$  to emphasize
    that all these (sign choice and higher order effects)
   cause  very minor modifications as is apparent when one
   compares Fig. 3 and Fig. 5.

\item The variation with $m_{top}(M_X)$ is quite marginal. An
example of how it affects the mixing near a maximal mixing point
is shown in Fig. 6.

\item Similarly the variation with $\chi_c$  near a maximal mixing
point is shown in Fig. 7.

\item The role of $m_c$ is essentially negligible since it is
$O(\epsilon^4)$ relative to $m_t$ and the equations are sensitive
only to their ratio.

\end{itemize}

\begin{figure}[h!]
\begin{center}
\epsfxsize15cm\epsffile{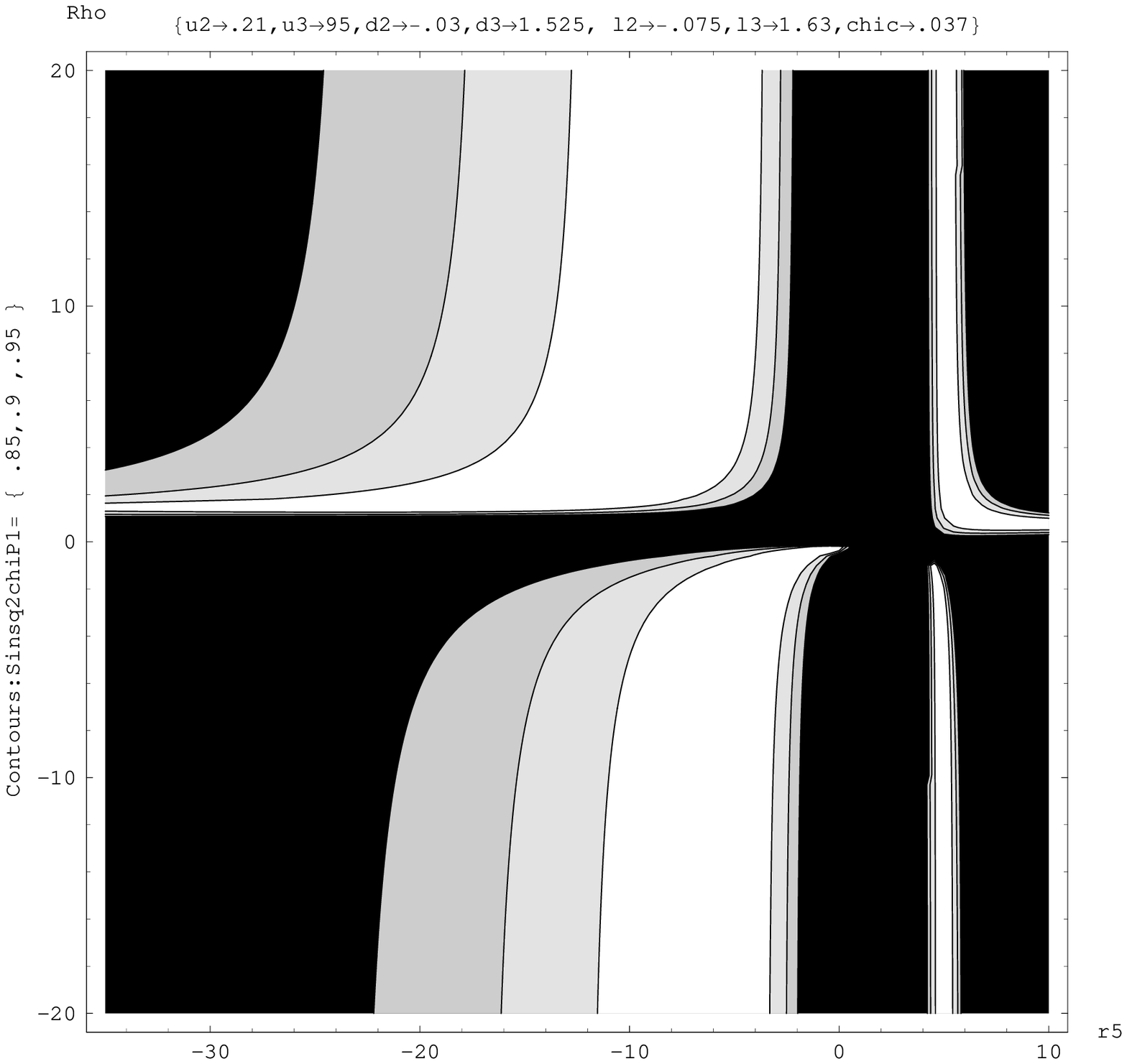}
 \caption {Contour plot of the exact solution for
 the  mixing parameter
 $Sin^2 2 \theta_{23} =Sin^2 2 \chi_P^{(1)}$ on the $r_5,\rho $
 plane,   at central values of the  charged fermion
 parameters(given on plot). The contours shown are at
 $Sin^2 2 \chi_P=.85,.9,.95$ with the shading
 becoming darker towards smaller values. The white region
 corresponds to $\sin^2 2\chi_P>0.95$.
 Note the  $\{sgn[d_2],sgn[l_2]\}=\{--\}$
 sign choice. This may be compared with Fig. 3.}
\end{center}
\end{figure}

\begin{figure}[h!]
\begin{center}
\epsfxsize15cm\epsffile{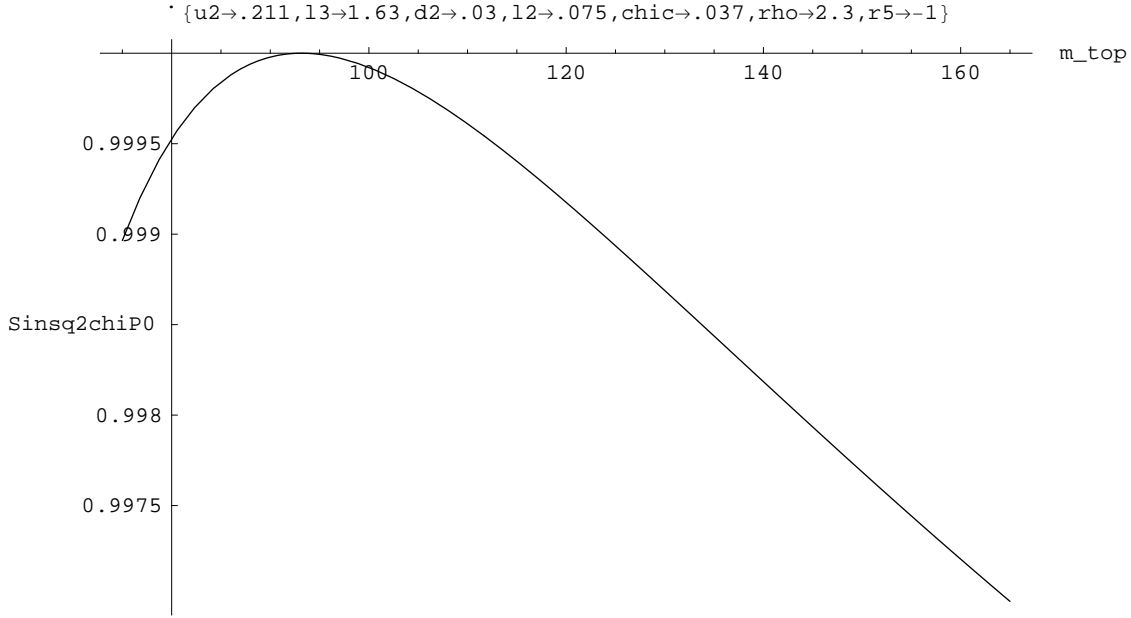}
 \caption{ Plot of the exact solution for the mixing parameter
 $Sin^2 2 \theta_{23} =Sin^2 2  \chi^{(0)}_P$,   vs
  $m_{top}(M_X)$ , with  sign choice $\{sgn[d_2],sgn[l_2]\}=\{++\}$}
\end{center}
\end{figure}

\begin{figure}[h!]
\begin{center}
\epsfxsize15cm\epsffile{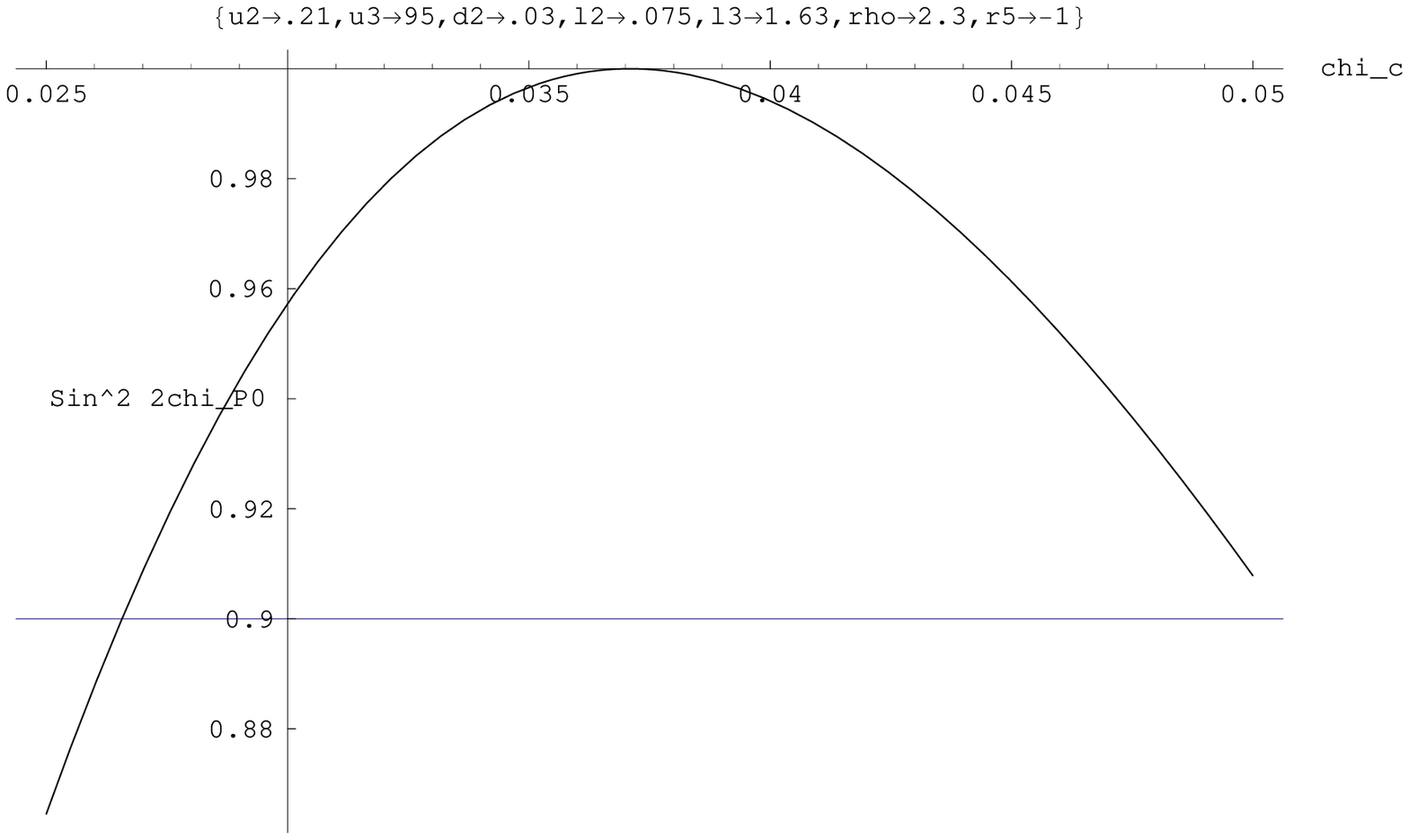}
 \caption{ Plot of the exact solution for  the mixing parameter
 $Sin^2 2 \theta_{23} =Sin^2 2  \chi^{(0)}_P$, to lowest
 order in $\epsilon$, vs
  $chi_c$  with  sign choice  $\{sgn[d_2],sgn[l_2]\}=\{++\}$}
\end{center}
\end{figure}

\section{Discussion, Conclusions and Outlook}

The work presented in this paper is based upon the very productive
insight \cite{bsv} that it is the   mixing in the 2-3 sector of
the Fermion mass matrices that is the true stable,   organizing
core of of the entire complex hierarchical fermion mass matrix.
Possibly due to the fact that the mixing angle $\theta^c_{23}$  in
the CKM matrix is much smaller than the Cabbibo angle
$\theta^c_{12}$  this realization has been slow to be digested by
the many workers fascinated by the the problem of finding a
non-trivial insight into the apparently finely crafted inner
workings of this wondrous ``primal artefact''.

Counter to the fact that   $\theta^c_{23}  <<\theta^c_{12}$  one
must consider the extreme tininess of the first generation masses
relative to those of the 2-3 generations, which ensures that we
must begin by understanding the smaller mixing angle. It was not
until the discovery of the tiniest mass (differences), via the
observation of neutrino oscillations, precisely in the range
expected from a seesaw between the Electro-Weak and GUT scales
that the deep inner connections between the physics of the Largest
(GUT), Smallest ($m_{\nu}$), and ``Everyday Electroweak "($M_W$)
mass scales emerged into plain view\cite{babpatwil,gmblm,blmdoom}.
The initial surprise at the natural explanations lurking in the
analysis of the simple 2-3 system and the apparently unique and
natural simplicity \cite{bsv} of the Type-II Seesaw, were followed
by further surprises at the demonstration that similar large
mixing could\cite{babmacesanu} also be obtained
(``sporadically''?) in some of the possible
 solutions of the  \emph{generic} BM\cite{babmoh}
fitting problem in Type I seesaw (or a combination of both seesaws).
  The price paid for this ``numerical discovery''
  was that it was based on  tuning the many
  unknown phases present in these models to achieve
the desired large neutrino mixing.

Unfortunately the BM program appears to
meet\cite{gmblm,blmdoom,bmsvkor,bmsv2} obstacles to its yield of
large  enough neutrino masses
 both in its Type I \cite{ckn,gmblm,blmdoom} and Type II
\cite{gohmohnasri,gmblm,blmdoom,bmsvkor,bmsv2} versions
 at least in the most desirable context of the MSGUT.
 Still this experience with the fitting problem
  provided the valuable lesson that the phase
  complications of the full
  complex CKM matrix problem would respect in large part the pattern
  discovered in the \emph{real } two and three generation
  fitting exercise\cite{babmacesanu} : which expectation
   is one of the operational assumptions of  our work.

The likely smallness of the  seesaw masses was noticed long before
our work  \cite{ckn,gohmohnasri} and was considered
 in \cite{gohmohnasri}  to be a motivation
 for constructing more elaborate GUT scale  Higgs structures to evade
 the tight constraints\cite{abmsv,ag2,gmblm,blmdoom}
 imposed  on seesaw coefficients  by the very simple
 yet fully functional  (and fully calculable\cite{ag1,bmsv,fuku04} !)
  structure of the   MSGUT. Our contribution was a systematic
  survey of the entire parameter space of the MSGUT
  to show that the   Type II contribution could not be saved by
   enhancement  at the special points
  of the MSGUT parameter space and that the same also went for
  Type I. These conclusions were based on the genericity of the
  values extracted from Type I,II fits available in the literature
  and need to be confirmed on a wider sample  generic fits or
  more satisfactorily by an analytic (perturbative) solution of
  the fitting problem  in the MSGUT\cite{nmsgutII}.
   We are personally reluctant to enthuse about the
  assumption of more complex AM  Higgs (i.e GUT scale SSB )
  structures. A ``shotgun '' programmatic  attitude towards
  decoding   the enigmas of unification seems questionable,
  because if there is no tight discipline restraining
  ``multiplication   of hypotheses''(representations)
   while speculating on extensions of established theories
then it is difficult, if not impossible, to know what one would be
testing,   if one ever could begin to do so : the wideness of the
space of extensions or any specific unavoidable implication of
theory. Since our knowledge about nature is sure to arrive only
incrementally and after hard struggle to define and test every
additional observable, it should be kept in mind that extensions
beyond those ``minimally'' necessary and forced upon us by the
inner logic of  Theory as we understand it in its essences -rather
than a patchwork hypothetical ``fix'' of its
 symptoms- should guide  our willingness to entertain new models:
  given  that the ``enveloping theory space '' of
  established Theory is almost  by definition  multiply
   infinite. If we seek to avoid the (in our view very fortunately
    found) tight  corners into which we are successful in driving the simplest -otherwise
 viable - model  by changing the model itself so completely that
 the entire previous simplifying and constraining
  understanding must be jettisoned then we foreclose any opportunity
   of hearing the ``tiny voice within''(the theory that is !)
   by  which nature may seek to inform us of the
    path to the decoding of these ultimate enigmas.

 The above pontifications may seem too sententious,
 pretentious or subjective, since one theorist's
 minimality is another  theorist's  monstrous pedantry!
   So we put it in more  concrete terms :
 changing the MSGUT's AM Higgs structure by adding  additional
$\mathbf{54}$'s or other Higgs multiplets will likely completely
destroy the GUT scale simplicity and calculability  achieved -
after tedious labour - in the current  formulation of
the\cite{ag1,abmsv,ag2,bmsv,fuku04}  MSGUT. We think it is
preferable to first allow the $\mathbf{120}$-plet arbitrarily
excluded by the BM-scenario (justifiably so at that early stage of
searching for the simplest viable model but now
 not an assumption worth protecting by
\emph{ad hoc }measures ):
which leaves the GUT scale SSB unchanged. In this
way one may hope to arrive at definite statements and
falsifiability about the simplest viable model so as to define the
next necessary(and falsifiable !) one. On the other hand extending
the GUT scale structure to avoid the conflict with the FM data
merely extends the catalog of non-falsifiable models without
 adding insight into the inner logic of the minimal
 theoretical structures adequate to encompass the data.

            Indeed, as shown in \cite{blmdoom} the constrained analysis
in the framework of the MSGUT points to the root of the problem
lying in the multiple contradictory constraints faced by  the
$\mathbf{\oot}$ Yukawa  couplings. A very simple \emph{calculable}
alternative, namely $\mathbf{10-120}$ domination of the charged
fermion sector coupled with very weak $\mathbf{\oot}$ couplings
that accentuate the Type I seesaw-- due to its inverse dependence
on these couplings through  the large mass of the  integrated out
heavy right handed   neutrinos -- emerges as an obvious and well
motivated resolution of the problem. Surprisingly just this
possibility  seems not to have been used to simplify the
\emph{prima facie }intractable nature of the fitting problem
involving both the two symmetric Yukawa matrices
 of the $\mathbf{10,120}$  \emph{and}  the antisymmetric couplings
 of the $\mathbf{120}$-plet. Indeed, guided by this insight, analysis
 of the scenario quickly yields  rather  novel insights into the
 structure of the fermion mass hierarchy. In particular,
 as shown by us in this paper,the mysterious  maximality
 ($Sin^2 \theta^{\nu}_{23} \sim 1.02 \pm .04$ \cite{strumviss})
 of the atmospheric mixing angle-- \emph{which has \emph{no} naturalness
  even in the  BM Type I and Type II seesaw
  fits}\cite{bsv,gohmoh,bert,babmacesanu}
   -- here emerges as a direct
  consequence of the most accurately known fermion
  mass data at the GUT scale and the well motivated
   structural pattern chosen for the theory
    due to its previous debacle\cite{blmdoom} !

  Usually such ``coincidences '' are explained by means
  of somewhat thinly motivated additional discrete
 symmetries or textures. Yet here Nature seems once again
 to hint that the solution is both simple and ``commonsensical'' and
 at the same time more deep and profound than anything arbitrary
 speculation might have rigged. The fermion spectra at $M_X$ seem
 seem to be fully compatible with the observed maximality of mixing
 without any assumption besides   allowing all
  $\mathbf{SO(10)}$ FM Higgs types
 and a pattern of dominance (and ``division of labour'')
 that assigns each Higgs a consistently bearable ``workload'' : all
serendipitously arranged so as to robustly  yield
 \emph{maximal} atmospheric
 mixing angle  fixed at a nearly  ``geometric '' value!

 Moreover the analysis has proved to be fully compatible with
  and robust under a systematic perturbation theory in which  the
  CKM angle $\theta^c_{23}$ is  considered  as a common small number
for structuring magnitudes of masses and angles
 (i.e universal Wolfenstein parameter) in the whole hierarchy,
 leaving behind magnitudes $O(1)$ or smaller to be determined order by
 order in the expansion of the fitting problem.
 An expansion of of the \emph{three } generation CP conserving
  case in this small parameter yields \cite{nmsgutII}
 a solution of the fitting problem that is based
 upon and respects the 2-generation analytic solution  found
 in this paper as its robust core. Besides the remarkable
   $\mathbf{b-\tau=s-\mu}$ unification constraint--which  strengthens
the rough $b-\tau$ equality    long canonical in  SUSY GUTs into
the  viable yet imminently falsifiable (i.e pending refinements in
the bottom quark mass measurement etc) $b-\tau$ pinning
 described above  we find  at $O(\epsilon^3)$
 an additional   constraint  between \emph{angles} :
 $\mathbf{\theta^c_{13}=\theta^c_{12}\theta^c_{23}}$,
 which is compatible with observation
 at least as regards  order of magnitude. Moreover since it
  arises at $O(\epsilon^3)$, by which order we expect CKM  CP
  violation to rear its head,   there is every reason to
  hope that this constraint will be modified to a more viable one
  involving the CKM phase   when the full 3 generation
  analysis is completed\cite{nmsgutIII}.
   Once the formalism of  generic fitting problem has been
   clarified it will be ready    for application in the Nu (realistic)
   MSGUT   (NMSGUT) whose  mass spectrum, zero mode Higgs  couplings,
and Baryon violating operators have been
calculated\cite{csask120} within the framework and conventions
of our decomposition\cite{ag1} of SO(10) group theory
to the maximal unitary
sub group $SU(4)\times SU(2)_L\times SU(2)_R$
(i.e the Pati-Salam group\cite{patisalam}).

Finally we note that the very nature of our \emph{ansatz} dictates
that the righthanded neutrino masses lie in the range below
$10^{12} GeV$ due to the smallness of the $\mathbf{\oot}$
couplings. Combined with the \emph{lower} bound\cite{davibarra}
that arises in the Leptogenisis models \cite{fukuyana} naturally
associated with SO(10) and the tight yet achievable pinning of
$m_b(M_X) $ to the more accurately known $m_\tau(M_X)$, we may
hope that our approach has added two significant stable elements
to the slowly but surely emerging picture of the intricate
interconnections and tightly interdependent structures  concealed
withinin the  Fermion mass Hierarchy.

We  earlier likened\cite{blmdoom} the circular linkages
 and mutual balance between the Large, Small and Geometric Mean
mass scales to the fabled cosmic serpent that swallows its own
tail : an ancient hermetic  symbol expressing the same sentiment of
wonder at the ``boundary less'' balance and intricate
 self sufficiency of the cosmic order that we seek to decipher.
 The wondrously patterned and enigmatic  Fermion Mass pattern
  is then perhaps the  supremely  fascinating and valuable
 diadem that the  $o\upsilon\rho\beta \acute{o}\upsilon\rho o\varsigma$
\footnote{\emph{Gr} : Tail-devourer} is fabled to wear.
 What was still
only the `spoor of a grail'\cite{ag2} now gleams -to our hopeful
eyes - with the iridescence of the \emph{ouroborotic
nagachudamani}\footnote{ \emph{Skr} : (Cosmic)
 Snake crestjewel/diadem}
which must -of course - also be  our true philosopher's stone !.

\section{Acknowledgements}

I am grateful to Sumit Garg for technical help, to Alejandra Melfo
for graphical inspiration and  to B.Bajc for correspondence.

\section{Note Added}

On receiving an advance version of this paper,
  B. Bajc  informed  me that in their latest work on the
$10-120$ system including complex couplings for the 2 generation
case they have also found restrictions on $m_b-m_\tau$ very
similar to those that form one of the central results of this
paper\cite{borutlate}.

 \end{document}